\documentclass[letterpaper, amsmath, amssym]{revtex4-2}

\usepackage[a4paper, total={6in, 9in}]{geometry}
\usepackage[T1]{fontenc}
\usepackage{titlesec}
\usepackage{array}
\usepackage{siunitx}
\usepackage[normalem]{ulem} 
\usepackage{xcolor}
\usepackage{booktabs} 
\usepackage{multirow} 
\usepackage{comment}

\usepackage{graphicx}
\usepackage[figurename=Figure, labelsep=period, font=scriptsize, labelfont=bf, format=plain]{caption}
\usepackage{placeins} 

\usepackage{bm} 
\usepackage{mathrsfs}
\usepackage{exscale}
\usepackage{easybmat}
\usepackage{mathtools}

\usepackage{hyperref} 
\usepackage{cleveref} 

\hypersetup{
    colorlinks=true,
    linkcolor=blue,     
    citecolor=blue,     
    urlcolor=blue       
}

\renewcommand\thesection{\arabic{section}}
\titleformat{\section}
  {\normalfont\bfseries\raggedright}{\thesection.}{1em}{}
\renewcommand\thesubsection{\thesection.\arabic{subsection}}
\titleformat{\subsection}
  {\normalfont\bfseries\raggedright}{\thesubsection}{1em}{}
\renewcommand\thesubsubsection{\thesubsection.\arabic{subsubsection}}
\titleformat{\subsubsection}
  {\normalfont\bfseries\raggedright}{\thesubsubsection}{1em}{}

\begin{document}

\title{An Integrated Lab on a CD Microfluidic Platform for High-Efficiency Blood Cell Separation and Passive Mixing}

\author{Reza Lotfi Navaei}\altaffiliation[Corresponding author ]{(rlotfina@buffalo.edu)}
\author{Haniyeh Tehrani}

\affiliation{\vspace{2ex} \makebox[\linewidth][c]{{\normalfont }\footnotesize{Dept. of Mechanical and Aerospace Engineering, University at Buffalo (SUNY), Buffalo, NY 14260-4400, USA}}
}

\begin{abstract}
Blood accounts for 7-8\% of total body weight, with an average adult containing 4.5 to 6 quarts. It delivers oxygen and nutrients to cells, removes waste products, supports immunity, and regulates body temperature. Comprising over 4,000 components, including plasma, platelets, erythrocytes, and leukocytes, blood presents a challenge for isolating specific cell types due to its heterogeneity. In autologous therapies, target cells may be as rare as one per million background cells. This study presents the development of a compact disc (CD)-based microfluidic device for high-throughput, label-free separation of blood components. The system integrates three main modules on a single disk: cell sorting, fluid control, and cell lysis. The initial module utilizes Pinched Flow Fractionation (PFF) to separate red blood cells, white blood cells, and platelets based on size, achieving 99.99\% efficiency. A capillary valve directs the sorted cells to the lysis module, where a chemical reagent is used to rupture cell membranes for downstream analysis. To enhance lysis efficiency, a micro mixer was incorporated using two zigzag geometries with internal barriers. The mixing performance of both designs was evaluated to determine optimal fluid interaction. This integrated, multifunctional CD platform offers a compact and effective solution for isolating and processing specific blood cells, with potential applications in diagnostics and personalized medicine.\\

\small{\textbf{Key words:} Micromixer, Lab on Chip, Cell Separation, Cell Lyse, Pinched Flow Fractionation }

\end{abstract}

\maketitle

\section{Introduction \label{sec:Intro}}
Cell separation is a crucial technique in biomedical science that facilitates the isolation of specific cell types from diverse populations for investigation, diagnostics, and therapeutic purposes \cite{daly_high_2023}. Effective cell separation is crucial in both clinical diagnostics, such as the isolation of circulating tumor cells (CTCs) from blood, and in research contexts for investigating stem cells or immune responses, as it ensures the precision and efficacy of subsequent processes \cite{ju_detection_2022, witek_cell_2020}. Microfluidics presents a tempting alternative to traditional cell separation methods by facilitating the precise manipulation of fluids and cells within microscale channels. These systems often utilize minimum reagents, provide high-throughput processing, and can be seamlessly connected with other analytical components for lab-on-a-chip applications. Cell separation in microfluidic devices can be accomplished through several mechanisms, categorized into passive and active approaches. Passive approaches utilize hydrodynamic forces, inertial effects, or deterministic lateral displacement to classify cells according to size, shape, or deformability. Active approaches employ external fields—such as electric (dielectrophoresis), magnetic, or sonic fields—to provide selective forces on target cells \cite{mansor_microfluidic_2025,10.1115/1.4067202}. These microfluidic techniques are especially beneficial owing to their scalability, delicate manipulation of cells, and capacity for real-time analysis.

Compact disc (CD) Microfluidics is a sub-field of microfluidics that deals with the behavior, precise control and manipulated of fluids in the micro-domain. This device embedding microfluidic channels and reservoirs into a CD-shaped plastic substrate uses centrifugal forces to move liquids. Using a motor, the whole platform spins and fluid manipulation is enabled \cite{liu2025rotary, rauf2025innovative,LEE2025103378}. For this system, a great range of fluidic operations has been developed including valving, decanting, calibration, mixing, metering, sample splitting, and separation. Often combined with analytical methods including optical imaging, absorbance and fluorescence spectroscopy, mass spectrometry, and mass spectrometry, these features make the centrifugal microfluidic platform a potent tool for clinical diagnostics and high-throughput drug screening. Two-point calibration of optode-based ion sensors, automated immunoassays, multiplexed screening assays, and cell-based assays are among CD-based centrifugal system applications. Various microscale separation methods have evolved in response to the need for effective cell separation, a vital first step in many biomedical studies. Advancement of actual point-of- care (POC) lab-on-a-chip (LOC) technologies depends on the development of effective microscale techniques allowing exact control over cell population distribution.

\begin{figure*}[!htbp]
    \centering
    \includegraphics[scale=0.4]{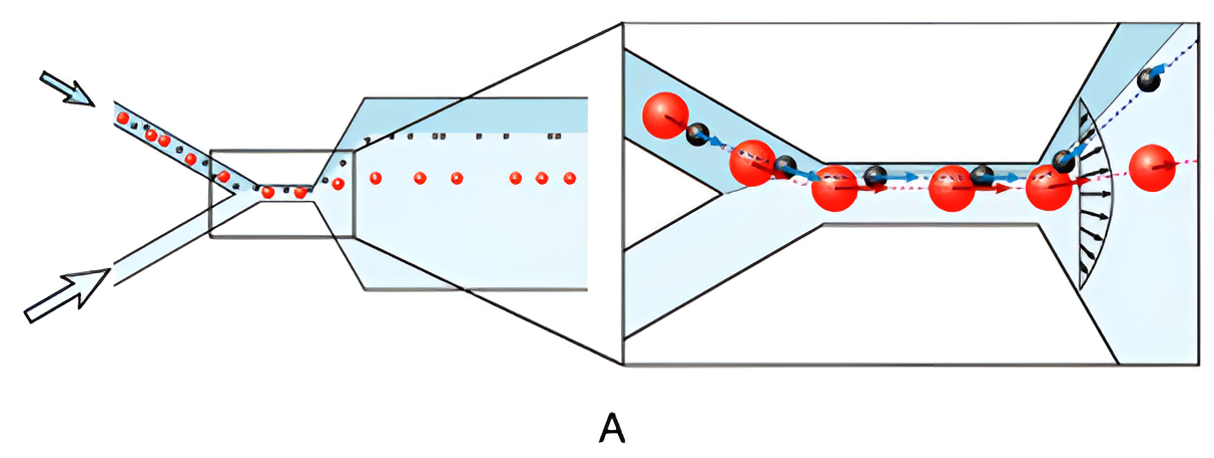}
    \captionsetup{justification=centering}
    \caption{A sample centrifugal cell separation system}
    \label{fig:1}
\end{figure*}

Lab-on-a-CD (LoaCD) systems are distinguished among microfluidic platforms for their innovative application of centrifugal force to facilitate fluid movement and execute intricate analytical tasks \cite{kordzadeh-kermani_low-cost_2025,Mitkees20042024}. These compact disk-shaped devices mimic a conventional compact disc and function by rotating on a motorized platform, where the resultant centrifugal forces propel fluids radially outward through microchannels and chambers (Figure \ref{fig:1}). The elimination of external pumps or valves streamlines system design, improves mobility, and lowers manufacturing expenses, rendering LoaCD especially appropriate for remote diagnostic applications\cite{dezhkam_lab---disk_2024, souderjani_numerical_2024}. In the last twenty years, LoaCD has been utilized in numerous biochemical experiments, sample preparation processes, and particularly in the separation of blood components and specific cell populations \cite{rondot_low-cost_2025}.

Cell separation in Lab-on-a-CD platforms use the consistent and adjustable characteristics of centrifugal forces to manage particles according to their physical attributes, including size, density, and sedimentation velocity \cite{miyazaki_biosensing_2020}. A prevalent method entails density-based separation, utilizing layered density gradients or sedimentation chambers to stratify cell populations, such as separate plasma from whole blood or extracting buffy coat layers that contain white blood cells \cite{moon_automated_2021,ZARASVAND2025100803}. Advanced LoaCD designs incorporate microstructures or obstacles to facilitate size-based filtration or deterministic lateral displacement directly on the spinning disc \cite{guevara-pantoja_microfluidic_2025}. These developments facilitate the efficient isolation of uncommon cells, such as circulating tumor cells (CTCs) or stem cells, within a compact, self-contained system. The versatility and automation capabilities of LoaCD platforms render them exceptionally appealing for consolidating many sample processing stages, such as separation, washing, lysis, and analysis, onto a single disk \cite{kutluk_integrated_2024, oconnell_integrated_2023}.

Recent research has shown substantial advancements in the implementation of Lab-on-CD systems for effective cell separation. Farahinia et al. \cite{farahinia_novel_2024} created a hybrid centrifugal microfluidic system that combines passive inertial sorting with magnetic separation to isolate circulating tumor cells (CTCs). Their approach attained approximately 93\% isolation efficiency by employing size-based bifurcation, succeeded by magnetic capture utilizing functionalized nanoparticles under optimum spin velocities and pulsatile regulation. A  completely automated Lab-on-a-Disk (LoaD) platform demonstrated the extraction of peripheral blood mononuclear cells (PBMCs) with minimal user involvement, highlighting the possibility for standardizing immune-cell separation processes was proposed by zhang et al. \cite{zhang_fully_2024}. Additionally, acoustic-assisted centrifugal methods have recently been investigated, providing label-free, high-throughput separation of microparticles and cells by integrating sound waves with disk rotation, hence improving selectivity and efficacy in microfluidic separation \cite{zaheri-ghannad_acoustic-assisted_2024}.

Lab-on-Disc (LoD) technology has swiftly progressed as a multifunctional platform for various biomedical applications, especially in diagnostics, by facilitating the effective control of different fluidic processes on a singular, compact disc \cite{kong_lab---cd_2016}. This capacity is essential for creating fully integrated "sample-to-answer" analysis systems that optimize intricate diagnostic operations. Zhang et al. \cite{zhang_fully_2024}. documented a LoD microstructure that can segregate blood cells from whole blood into separate reservoirs in about one second, attaining a red blood cell separation efficiency of up to 99\%. Park et al. created a completely automated disc platform that effectively processed 5 mL of blood for the high-purity separation of circulating tumor cells (CTCs), which are essential for early cancer detection and monitoring \cite{plouffe_perspective_2014}. Moreover, Lee et al. \cite{lee_fully_2009} exhibited a fully integrated ELISA on a disc for the direct detection of Hepatitis B virus (HBV) biomarkers from whole blood, highlighting the platform's capability for sophisticated immunoassay automation.

Passive methods, on the other hand, depend entirely on channel design and hydrodynamic forces. For example, the Pinched Flow Fractionation (PFF) method separates cells based on size and hydrodynamic effects. Several parameters influence separation efficiency in PFF, including the width and length of the pinched segment, the width of the broadened segment, the angle between these segments, the sample-to-buffer flow rate ratio, and the parabolic velocity profile (Poiseuille flow). This method offers several advantages, such as ease of design and fabrication, and the ability to separate particles of various sizes.

This method proposed by Yamada et al. \cite{yamada2004pinched}. Microchip that they had designed was capable to separate micro particles with diameters of 15, 30 microns by 20 micro liters per hour and more than 90\% efficiency. Morijiri et al. \cite{morijiri2011sedimentation}, fabricated a microchip based on this method that it can separate particles with diameters of 3 and 5 micron by 3090 micro liter per hour fractionation rate. A simple and efficient device for density-based particle sorting is in high demand for the purification of specific cells, bacterium, or environmental particles for medical, biochemical, and industrial applications \cite{bozorgpour2023computational, bozorgpour2023exploring, jafari2025metamaterialbistablevibrationabsorbers}. For example, this idea can serve as an additional approach to addressing neck injuries caused by vibrations and forces, a potential area investigated by Afsharfard et al. \cite{afsharfard2023modifying}.

Mixing is another critical operation in biological sample processing. At the macroscale, mixing is typically achieved through turbulent flows using mechanical elements such as blades or stirrers. However, at the microscale, generating turbulence is impractical; instead, mixing predominantly relies on molecular diffusion \cite{dolatyari2025thermal}. Microfluidic mixers are generally classified into two categories: passive and active. Passive mixers operate without moving parts, relying on pressure gradients to drive fluids through specially designed microchannel geometries. These systems typically function at low Reynolds numbers and are optimized to increase fluid contact surface area, thereby enhancing diffusion. Due to spatial limitations, passive mixers must be engineered to maximize interaction length within a confined footprint. For instance, Wang et al. \cite{wang2013numerical} studied a tree-like, T-shaped microchannel both numerically and experimentally. They found that increasing the number of T-branches enhanced mixing performance at low Reynolds numbers. Similarly, Moheb et al. \cite{ait2012numerical} compared T-shaped and cross-shaped microchannels, concluding that cross-shaped channels not only offer better mixing efficiency at low flow rates but also result in lower pressure losses. Yang et al. \cite{yang2015high} explored a 3D micromixer based on Tesla geometry for biological applications. In addition to the design and fabrication of the device, they analyzed the flow behavior and demonstrated that it operates efficiently across a wide Reynolds number range (0.1 to 100), making it effective for applications like antibody mixing in cancer cell detection \cite{khaniki2024brain}. 

On the other hand, active microfluidic mixers involve moving components or particles, driven by external stimuli. A common example is the use of magnetic particles, which are actuated via externally applied magnetic fields. Other methods involve manipulating fluids using thermal gradients, electric fields, or pressure variations \cite{jafari2022conceptual}. These active systems play a crucial role in microfluidic system efficiency and are increasingly important in advanced microfluidic applications. One great example is the new active microfluidic droplet generator proposed by Amiri et al. \cite{amiri_shear-thinning_2021}. This study simulates the formation of non-Newtonian, shear-thinning droplets in a microfluidic T-junction under the influence of an external electric field, using a level set numerical method. It investigates how key parameters—such as inlet velocities, droplet-phase concentration, and applied voltage—affect droplet size and formation time.

Micromixers are also widely used to perform a variety of chemical operations, including cell lysis, a crucial step in many bioprocesses. One of the most significant  applications of cell lysis is DNA extraction from cells \cite{khaniki2024vision, abbasi2018modeling}. This process involves breaking down cell walls and isolating DNA from other cellular components. The procedure can be simplified using specific reagents. For example, the membranes of cultured animal cells can be easily disrupted with detergents such as sodium dodecyl sulfate (SDS), which releases intracellular contents. However, some cells possess more rigid cell walls, requiring a combination of enzymatic and chemical treatments for effective lysis. The lysozyme enzyme degrades polysaccharide components of the cell wall, while EDTA helps weaken the wall by chelating magnesium ions, making it more susceptible to destruction. Following cell wall breakdown, detergents are used again to lyse the cell membrane. To isolate DNA, it is necessary to remove all remaining cellular components. Detergents such as SCTAB are effective for this purpose, as they bind to nucleic acids and aid in their separation from other molecules. As evident from this process, effective mixing is a critical factor in cell lysis. The fluid carrying the cells must be thoroughly mixed with the lysis reagents to ensure complete and uniform lysis. A notable example of a microfluidic system designed for this purpose is presented by Ashkani et al. \cite{ashkani2024enhancing}, who developed a micromixer utilizing square-wave microchannels to combine the lysis buffer with cell-containing samples. They experimentally investigated the rate of cell lysis across microchannels of varying lengths and also examined how different buffer compositions influenced the efficiency of the lysis process.

\begin{figure*}[!htbp]
    \centering
    \includegraphics[width=15.5cm, height=9.5cm]{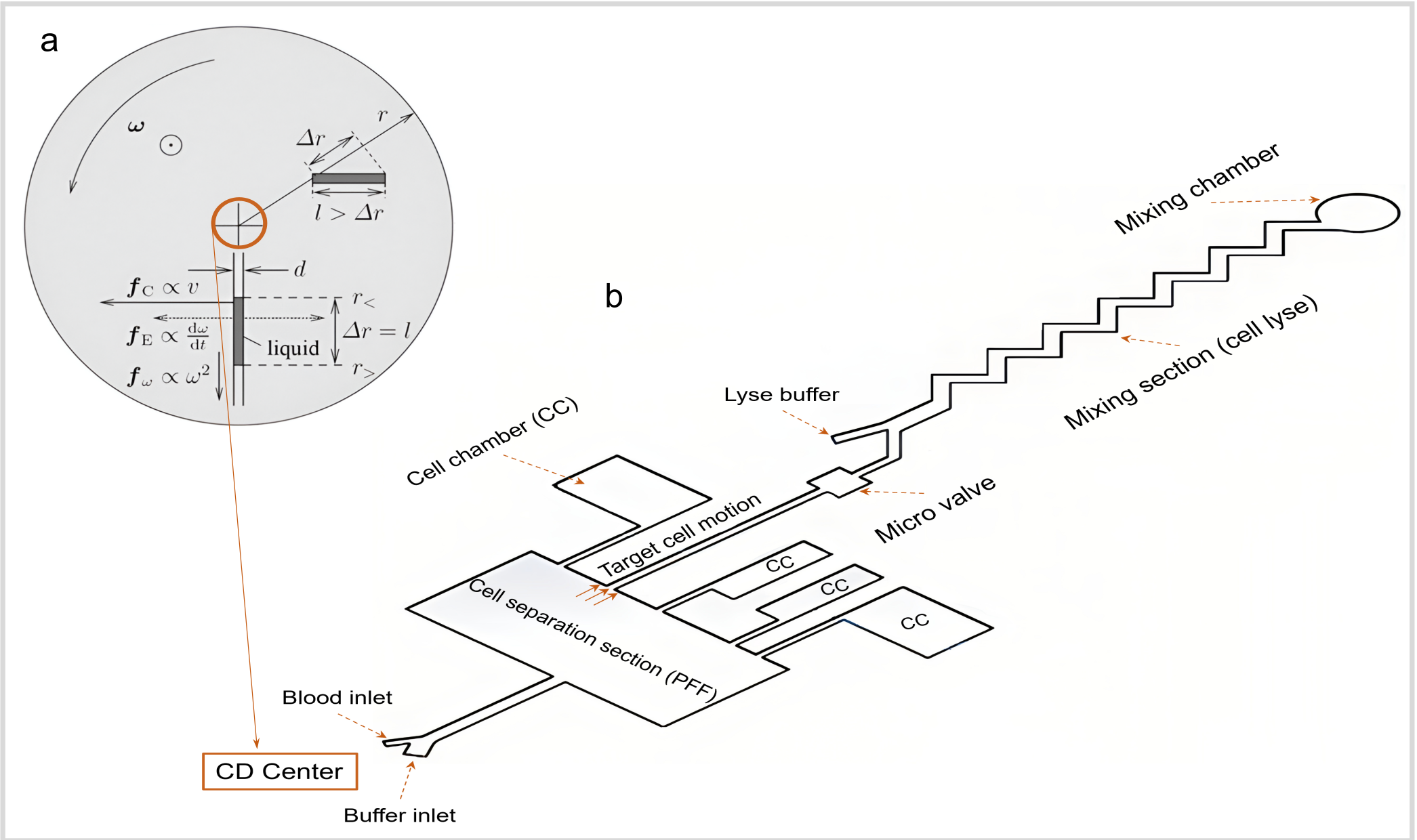}
    \captionsetup{justification=centering}
    \caption{Design shematic of Lab on CD microfluidic system for cell separation and mixing}
    \label{fig:2}
\end{figure*}

Despite the notable advancements and varied applications of Lab-on-Disc technology in cell separation, a considerable challenges still in the development of fully integrated devices that incorporate numerous intricate operations within a single disk. Although existing systems can execute several processes, the realization of smooth, multi-step workflows on a single, self-sufficient disc, with reduced reliance on external components and operator intervention, is a continual engineering challenge \cite{xiao_fully_2024}. Numerous current platforms continue to depend on external fluidic control or necessitate off-chip computation for specific functions, hence constraining their complete autonomy and portability \cite{michael_challenges_2016, afsharfard2014modeling}. In this study, we present microfluidic systems designed for size- and density-based particle separation, utilizing sedimentation effects and a microscale hydrodynamic technique known as Pinched-Flow Fractionation (PFF). Also, This study seeks to examine the essential design factors and technological innovations needed to bridge this gap, concentrating on the creation of a fully integrated Lab-on-Disc device that can perform all requisite cell separation and analysis functions within a single disk as shown in~\Cref{fig:2}.

\FloatBarrier
\section{Methodology}
The fluid within the pinched flow fractionation (PFF) device is modeled as a Newtonian and incompressible fluid, consistent with the behavior of aqueous solutions typically used in microfluidic applications. Under these assumptions, the fluid motion is governed by the incompressible Navier–Stokes equations, which describe the conservation of momentum, coupled with the continuity equation representing the conservation of mass. These equations provide a fundamental framework for capturing the laminar, pressure-driven flow characteristics observed in the microchannel, where viscous effects dominate and inertial contributions are minimal due to the low Reynolds number regime {Eq.\eqref{eq:navier_stokes}}:
\begin{equation}
    \rho_f (\partial_t + \vec{u} \cdot \nabla) \vec{u} = -\nabla P + \mu \nabla^2 \vec{u} + F
    \label{eq:navier_stokes}
\end{equation}
\begin{equation}
    F = F_{ce} + F_{co} = \rho (r |\omega|^2 - 2\omega \times V)
    \label{eq:forces}
\end{equation}

where $\rho_f$ is the density of the fluid, $\vec{u}$ is the fluid velocity, $P$ is the pressure, $\mu$ is the dynamic viscosity (the kinematic viscosity $\nu$ is $\mu/\rho$), $F$ is the sum of all body forces, $r$ is radial position vector, $\omega$ is angular velocity vector, $F_{ce}$ is the centrifugal force, and $F_{co}$ is the Coriolis force.\\
Mass conservation implies the continuity equation:
\begin{equation}
 \nabla \vec{u}=0
\end{equation}

\subsection{Governing equations for mixing:}      
The Reynolds number is a fundamental dimensionless parameter used to characterize the flow regime within a fluid system by quantifying the relative influence of inertial forces to viscous forces. It serves as a key indicator in determining whether the flow is laminar, transitional, or turbulent, and is especially critical in microscale fluid dynamics, where it helps predict flow behavior and informs the appropriate selection of modeling approaches and assumptions. 

\begin{equation}
 Re=\frac{\rho u d}{\mu} 
\end{equation}

In which $\rho$ is fluid density, u is the fluid flow velocity, d is the characteristic length and $\mu$ is the fluid dynamics viscosity. In microchannel flows, the Reynolds number is a critical parameter in determining the nature of the fluid motion. For Reynolds numbers typically below 2000 to 2500, the flow remains in the laminar regime, characterized by smooth, orderly fluid motion with minimal mixing due to turbulence. In this study, the Reynolds number remains well within this range, justifying the assumption of laminar flow throughout the domain. To describe the conservation of mass and momentum for an incompressible and Newtonian fluid under these conditions, the governing equations consist of the continuity equation and the incompressible Navier–Stokes equations, which are expressed as follows:

\begin{equation}
\boldsymbol{\rho} \frac{\partial \boldsymbol{u}}{\partial t} - \nabla \cdot [-P\mathbf{I} + \eta (\nabla \boldsymbol{u} + (\nabla \boldsymbol{u})^T)] + \rho \boldsymbol{u} \cdot \nabla \boldsymbol{u} = \boldsymbol{F}
\end{equation}

In which u is the velocity vector, $\rho$ is the fluid pressure, $\mathbf{I}$ is unity matrix and F is the volume force affecting the fluid. In addition to the flow field, the transport of solutes within the fluid is governed by the convection–diffusion equation, which accounts for both advective transport due to fluid motion and molecular diffusion driven by concentration gradients. This equation provides a comprehensive description of the spatiotemporal evolution of solute concentration within the microchannel and is essential for accurately modeling mixing processes and species distribution in the system.

\begin{equation}
\frac{\partial C}{\partial t} + \nabla \cdot (-D\nabla C) = R - u\nabla C
\end{equation}

In which C is the concentration, D is the diffusion constant and R is the reaction rate. To quantitatively evaluate the mixing performance within the microfluidic device, the mixing efficiency was calculated using the standard deviation of the solute concentration distribution at the outlet cross-section. The mixing index, is defined as:

\begin{equation}
\sigma = \sqrt{\frac{1}{N}\sum_{i=1}^{N} (c_i - \overline{c}_m)^2}
\end{equation}

where N is the number of points from a cross section from which a concentration sample is taken, $c_i$ is the concentration of the sample i and $c_m$ is the optimum concentration we wanted to achieve, then:

\begin{equation}
M = 1 - \sqrt{\frac{\sigma^2}{\sigma^2_{max}}}
\end{equation}

In which $\sigma_{max}$ is the condition where there has been no mixing. We call M, the mixing index.  The mixing index is a dimensionless metric ranging between 0 and 1, where a value of 0 indicates no mixing (fully segregated fluids), and a value approaching 1 indicates near-perfect mixing, characterized by a uniform concentration profile across the cross-section. This index provides a normalized measure of mixing efficiency that facilitates direct comparison between different mixer designs or operating conditions.

\FloatBarrier
\section{Results}
In this study, we modeled and simulated the separation of blood cells using a microfluidic device designed based on the principles of pinched flow fractionation (PFF). To accurately represent the major cellular components of blood, we utilized three types of particles with diameters corresponding to white blood cells (WBCs), red blood cells (RBCs), and platelets. Specifically, the diameters of the particles were set to 15 µm, 8 µm, and 1.8 µm for WBCs, RBCs, and platelets, respectively. These values are consistent with the physiological size ranges observed in actual human blood components, ensuring that the simulations closely resemble real-world conditions.

To simulate the effects of rotational motion, which plays a crucial role in enhancing cell separation in rotating microfluidic systems, we incorporated both centrifugal and Coriolis forces as body forces in our numerical model. The centrifugal force acts radially outward from the axis of rotation and varies in magnitude based on particle density and radial position, thereby influencing the lateral migration of cells. Meanwhile, the Coriolis force, which arises due to the rotating reference frame, acts perpendicular to the particle velocity and rotation axis, further altering particle trajectories in a size- and velocity-dependent manner. The combined influence of these forces leads to distinct particle focusing and separation behaviors, which are critical for effective sorting of the different blood cell types.

To identify the most effective geometry for cell separation, we systematically simulated multiple microchannel designs with varying structural features, such as channel width, curvature, and the configuration of inlet and outlet branches. These geometries were assessed based on their ability to induce size-dependent focusing of particles and direct them toward distinct outlet regions. Through this iterative design and evaluation process, we ultimately determined an optimal channel configuration that maximized separation resolution while minimizing particle loss or crossover between outlet streams.

In constructing the simulation environment, we defined key physical parameters, including the densities and diameters of the particles, as well as the viscosity and flow rate of the carrier fluid. The simulations were carried out using a structured mesh adapted to the complex geometry of the microfluidic channel. The mesh generation process ensured sufficient resolution in critical regions—such as near channel walls and bifurcations—to capture the detailed hydrodynamic behavior of the particles. Boundary conditions were carefully specified to replicate realistic flow conditions: velocity inlets were assigned to the fluid and particle streams, while pressure-outlet conditions were applied to the multiple outlets. A no-slip condition was enforced at all channel walls.

\begin{figure*}[!htbp]
    \centering
    \includegraphics[width=15.5cm, height=9.5cm]{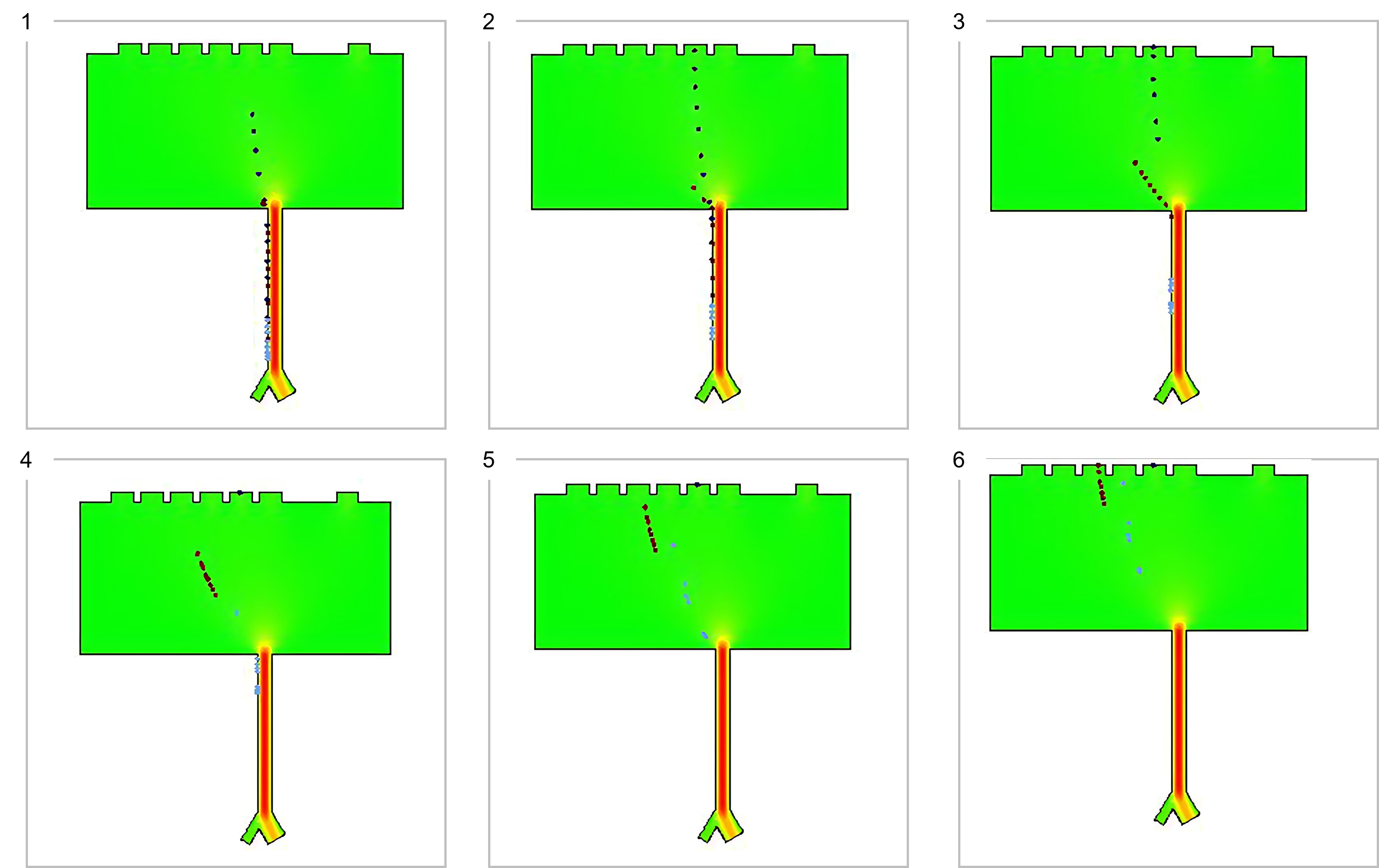}
    \captionsetup{justification=centering}
    \caption{Schematic of cell separation system and particle tracing for different sizes at: (1) t=0.85 s, (2) t=1.75 s, (3) t=2.6 s, (4) t=3.5 s, (5) t=4.05 s, and (6) t=5.5 s}
    \label{fig:3}
\end{figure*}

The particle-laden flow was then subjected to rotation at a fixed angular velocity of 1000 revolutions per minute (rpm), a value selected to provide a strong yet manageable centrifugal environment within the microfluidic system. Under these conditions, we tracked the motion of particles over time, focusing particularly on their trajectory development and final outlet destinations. The simulation results, depicted in Figure \ref{fig:3} insets, illustrate the temporal evolution of particle positions within the rotating microchannel.

Analysis of the simulation data revealed that the applied body forces effectively guided the particles along distinct streamlines according to their size. Larger particles, such as WBC analogs, experienced stronger centrifugal and Coriolis effects, which caused them to migrate laterally toward the outer regions of the channel and exit through the outlets positioned closer to the outer wall. Medium-sized particles (RBC analogs) followed intermediate paths, while the smallest particles (platelet analogs) remained near the inner wall and were collected through outlets aligned with the inlet sidewall streamlines. This size-based deflection behavior confirms the successful realization of particle separation in the rotating frame.

The observed trajectories and outlet distributions indicate that the separation process achieved a near-ideal efficiency. Each particle type was directed to a distinct outlet region with minimal overlap, demonstrating effective discrimination based solely on particle size. Moreover, particles with identical diameters consistently followed the same streamline paths and exited through the same outlet, confirming the high repeatability and reliability of the separation mechanism. Based on these findings, we conclude that the optimized PFF geometry, in conjunction with the applied rotational body forces, enables highly efficient, label-free separation of blood cell analogs, with a separation efficiency approaching 100 percent under the given operating conditions.

To investigate the hydrodynamic behavior and mixing efficiency within the microfluidic mixer, a numerical simulation was performed based on the governing equations for laminar flow and species transport. Given the characteristic dimensions of the microchannel and the flow conditions, the Reynolds number was calculated to fall within the range of 3 to 20. These values confirm that the fluid motion remains within the laminar regime, where viscous forces dominate and inertial effects are negligible. Under such low Reynolds number conditions, mixing by convection is limited, and molecular diffusion becomes the dominant mechanism for mass transport between fluid streams.

To accurately model this process, the flow field was solved using the Navier-Stokes equations for incompressible laminar flow, and the transport of solutes was captured by solving the convection–diffusion equation. This coupled approach enabled detailed simulation of fluid dynamics and concentration evolution throughout the microchannel.

A high-resolution computational mesh was generated over the fluidic domain to resolve fine flow structures and concentration gradients, particularly in regions with geometric complexity or sharp concentration interfaces. The mesh employed an element size ranging from 0.05 µm to 14.9 µm, ensuring both accuracy and numerical stability. A maximum element growth rate of 1.05 was used to maintain smooth transitions between element sizes, while a curvature refinement factor of 0.02 was applied to ensure proper resolution of curved channel features and embedded structures.

Rotational effects were included in the model by incorporating body forces representing centrifugal and Coriolis accelerations within the momentum equations. These volumetric forces arise due to the rotation of the platform and significantly influence the lateral migration of fluid elements and particles. The center of rotation was set at a distance of 2.5 cm from the center of the mixing section, a value that determines the strength of the centrifugal component and influences the efficiency of fluid manipulation within the microchannel.

The simulation modeled two separate fluid streams entering the mixer. The first stream was assumed to contain cells suspended in plasma with an initial solute concentration of 1 mol/m³, while the second stream consisted of an aqueous buffer containing the same solute concentration, representing a salt solution used for cell lysis. The interaction and mixing of these two streams were investigated under the influence of flow and rotational forces.

\begin{figure*}[!htbp]
    \centering
    \includegraphics[width=15.5cm, height=9.5cm]{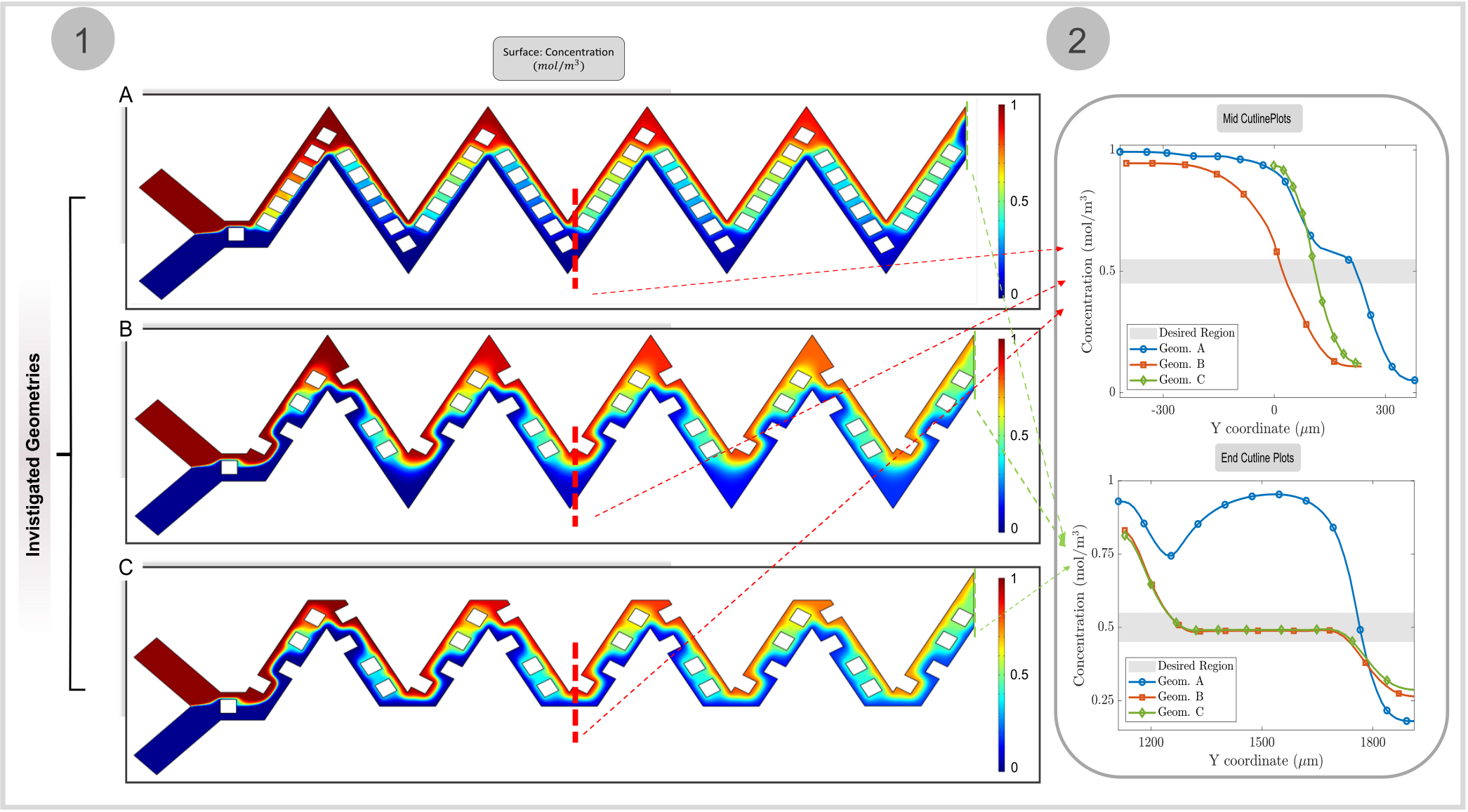}
    \captionsetup{justification=centering}
    \caption{Comparison of 2 different geometry for mixing in 1000 rpm}
    \label{fig: 4}
\end{figure*}

To enhance mixing, two different microchannel geometries featuring embedded barriers within a zigzag channel layout were considered. In the first geometry, the embedded barriers were placed along the centerline of each zigzag segment. In the second configuration, a staggered layout was employed: the first barrier was placed along the right wall at the start of the first segment, the second along the left wall at the beginning of the second segment, while the remaining barriers were centered as in the first design. This asymmetry was intended to disrupt flow symmetries and promote more effective mixing by inducing chaotic advection and enhancing interfacial area between the two streams as can be seen in Figure \ref{fig: 4}.

Mixing performance was assessed by evaluating the concentration field at the channel outlet. The results presented in Figure \ref{fig: 4}. A comparison of the two designs revealed that the second, asymmetrically patterned geometry led to more uniform solute distributions, indicating superior mixing efficiency. This result is visualized through concentration contours, which show significantly reduced gradients and a more homogeneous concentration field at the outlet of the second design. Consequently, the second geometry was selected for all subsequent simulations.

To further investigate the mixing dynamics within the selected microchannel geometry, a detailed set of simulations was performed under a rotational speed of 1000 revolutions per minute (rpm). The objective of this analysis was to assess how the mixing process evolved spatially along the channel length, from the initial point of fluid introduction to the final outlet region. The system was initialized with two distinct fluid streams entering the mixer through separate inlets, each carrying an identical solute concentration but no prior mixing. As the fluids progressed through the channel, they experienced the combined effects of laminar flow, geometric perturbations introduced by embedded barriers, and rotational body forces.

\begin{figure*}[!htbp]
    \centering
    \includegraphics[width=15.5cm, height=9.5cm]{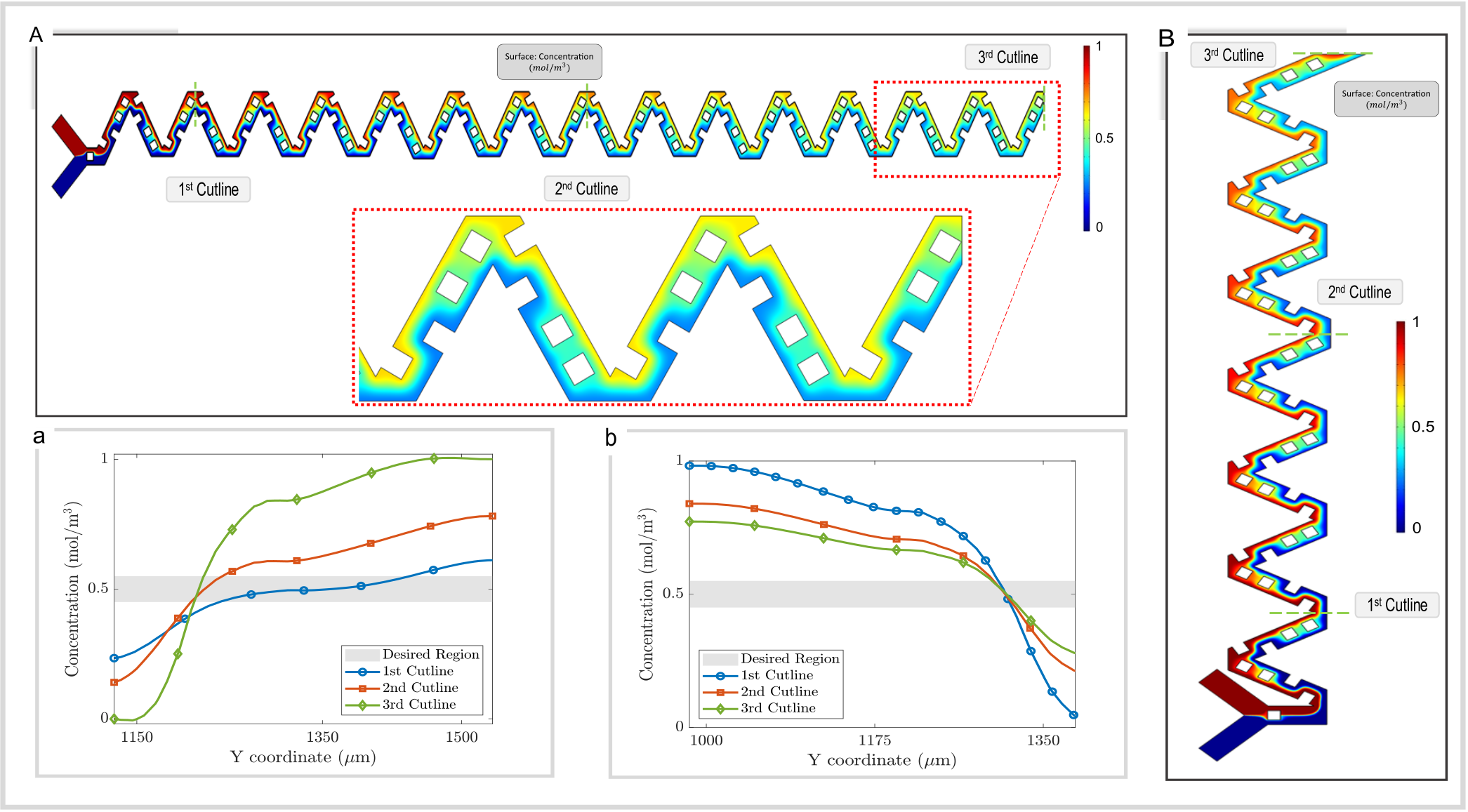}
        \captionsetup{justification=centering}
    \caption{Diagram of concentration changes in 3 mentioned cross sections for 1000 rpm}
    \label{fig:5}
\end{figure*}

The mixing progression was evaluated by analyzing solute concentration distributions at multiple positions along the channel as shown in Figure \ref{fig: 4}. Specifically, three cross-sectional planes were defined: one near the channel inlet, one at the midpoint, and one immediately upstream of the outlet. At each of these locations, both two-dimensional concentration contour maps and one-dimensional concentration profiles were extracted and analyzed. These cross-sectional concentration fields revealed a clear transition from sharp concentration gradients at the inlet—where the fluids remained largely unmixed—to nearly uniform solute distribution at the outlet, indicating successful mixing as shown in Figure \ref{fig:5}.

At the inlet, the two fluid streams maintained distinct boundaries with minimal interfacial diffusion, consistent with the low Reynolds number regime and short residence time. By the midpoint, lateral diffusion was noticeably enhanced, aided by the flow disturbances introduced by the micro-barriers and the lateral displacement induced by Coriolis and centrifugal effects. At the outlet, the concentration profiles flattened significantly, demonstrating that the solute had become well-dispersed across the channel width, resulting in a homogenous mixture. This progression confirmed that the designed channel configuration effectively facilitates mixing along the length of the device, even in the absence of turbulence or external actuators.

To evaluate the influence of rotational speed on mixing performance, additional simulations were conducted at higher angular velocities of 1500 rpm and 2000 rpm. These simulations were intended to examine how increased body forces affect the mixing process under otherwise identical flow conditions. The results revealed a clear trend: higher rotational speeds intensified the centrifugal and Coriolis forces acting on the fluid, which in turn enhanced the transverse motion of fluid layers and amplified the interfacial deformation between the two streams.

This increase in transverse mixing led to more rapid homogenization of the solute, as evidenced by faster convergence of concentration profiles across the channel width and earlier onset of uniform mixing along the channel length. At 1500 rpm, the interface between the two fluids became more distorted compared to the 1000 rpm case, resulting in faster diffusion across the contact plane. At 2000 rpm, the rotational forces were even more pronounced, causing strong lateral displacement and interleaving of fluid elements. This dynamic promoted high interfacial area exposure and efficient mixing within a shorter axial distance.

Collectively, the findings from all three rotational conditions consistently demonstrated the effectiveness of the selected geometry and barrier arrangement in achieving high-quality passive mixing. The simulations confirmed that mixing efficiency can be significantly enhanced by increasing the angular velocity, offering a controllable parameter for tuning performance in centrifugal microfluidic systems. The results highlight the critical role of rotational body forces in accelerating mixing under low Reynolds number conditions and support the use of rotational speed as a design and optimization parameter in the development of efficient microfluidic mixers.
\FloatBarrier
\section{Conclusions}
In this work, we have successfully developed and numerically validated a compact, multifunctional microfluidic platform designed for integration onto a single rotating disc substrate. The device is capable of performing multiple key operations sequentially, including cell separation, fluid routing, and on-chip mixing, all within a single, miniaturized system. The core design objective was to create a high-throughput, label-free, and scalable platform suitable for potential use in point-of-care diagnostic applications, where sample processing speed, automation, and device compactness are of critical importance. The initial module of the device focuses on the separation of blood components using a size-based pinched flow fractionation (PFF) strategy. This module allows for the continuous and simultaneous isolation of red blood cells (RBCs), white blood cells (WBCs), and platelets directly from whole blood, without the need for external labeling agents or biochemical tags. By exploiting the intrinsic size differences among these cell types—approximately 15 µm for WBCs, 8 µm for RBCs, and 1.8 µm for platelets—the device effectively directs each population along distinct flow paths within the microchannel, ultimately guiding them to designated outlet ports. The application of rotational forces, including both centrifugal and Coriolis components, enhances the separation resolution by introducing controlled lateral migration, which further amplifies the size-dependent deflection of particles. Simulation results confirmed a separation efficiency of approximately 99.9percent, indicating near-complete discrimination among the three primary blood cell types. Two distinct mixer geometries were evaluated to identify an optimal configuration for maximizing mixing performance. The first geometry featured symmetrically placed barriers along the centerline of the zigzag segments, while the second employed a staggered layout with alternating barrier placements along the sidewalls and centerline. Quantitative and qualitative analysis of solute concentration distributions across multiple cross-sections demonstrated that the staggered configuration resulted in significantly improved mixing uniformity. As a result, the second geometry was selected for all subsequent simulations and considered the most effective design for integration into the final platform. Taken together, the compact and highly integrated nature of this centrifugal microfluidic device underscores its potential for next-generation point-of-care applications. By combining high-efficiency, label-free cell separation with rapid on-chip mixing for downstream biochemical processing, this platform addresses several key challenges in diagnostic microfluidics, including miniaturization, automation, and multifunctionality.\\

\section*{References}
\bibliographystyle{IEEEtran}
\bibliography{Ref}

\end{document}